\documentstyle[sprocl]{article}
\newread\epsffilein    
\newif\ifepsffileok    
\newif\ifepsfbbfound   
\newif\ifepsfverbose   
\newif\ifepsfdraft     
\newdimen\epsfxsize    
\newdimen\epsfysize    
\newdimen\epsftsize    
\newdimen\epsfrsize    
\newdimen\epsftmp      
\newdimen\pspoints     
\def\epsfbox#1{\global\def\epsfllx{72}\global\def\epsflly{72}%
   \global\def\epsfurx{540}\global\def\epsfury{720}%
   \def\lbracket{[}\def\testit{#1}\ifx\testit\lbracket
   \let\next=\epsfgetlitbb\else\let\next=\epsfnormal\fi\next{#1}}%
\def\epsfgetlitbb#1#2 #3 #4 #5]#6{\epsfgrab #2 #3 #4 #5 .\\%
   \epsfsetgraph{#6}}%
\def\epsfnormal#1{\epsfgetbb{#1}\epsfsetgraph{#1}}%
\def\epsfgetbb#1{%
\openin\epsffilein=#1
\ifeof\epsffilein\errmessage{I couldn't open #1, will ignore it}\else
   {\epsffileoktrue \chardef\other=12
    \def\do##1{\catcode`##1=\other}\dospecials \catcode`\ =10
    \loop
       \read\epsffilein to \epsffileline
       \ifeof\epsffilein\epsffileokfalse\else
          \expandafter\epsfaux\epsffileline:. \\%
       \fi
   \ifepsffileok\repeat
   \ifepsfbbfound\else
    \ifepsfverbose\message{No bounding box comment in #1; using defaults}\fi\fi
   }\closein\epsffilein\fi}%
\def\epsfclipoff{\def\epsfclipstring{\ifepsfdraft\space clip\fi}}%
\epsfclipoff
\def\epsfsetgraph#1{%
   \epsfrsize=\epsfury\pspoints
   \advance\epsfrsize by-\epsflly\pspoints
   \epsftsize=\epsfurx\pspoints
   \advance\epsftsize by-\epsfllx\pspoints
   \epsfxsize\epsfsize\epsftsize\epsfrsize
   \ifnum\epsfxsize=0 \ifnum\epsfysize=0
      \epsfxsize=\epsftsize \epsfysize=\epsfrsize
      \epsfrsize=0pt
     \else\epsftmp=\epsftsize \divide\epsftmp\epsfrsize
       \epsfxsize=\epsfysize \multiply\epsfxsize\epsftmp
       \multiply\epsftmp\epsfrsize \advance\epsftsize-\epsftmp
       \epsftmp=\epsfysize
       \loop \advance\epsftsize\epsftsize \divide\epsftmp 2
       \ifnum\epsftmp>0
          \ifnum\epsftsize<\epsfrsize\else
             \advance\epsftsize-\epsfrsize \advance\epsfxsize\epsftmp \fi
       \repeat
       \epsfrsize=0pt
     \fi
   \else \ifnum\epsfysize=0
     \epsftmp=\epsfrsize \divide\epsftmp\epsftsize
     \epsfysize=\epsfxsize \multiply\epsfysize\epsftmp   
     \multiply\epsftmp\epsftsize \advance\epsfrsize-\epsftmp
     \epsftmp=\epsfxsize
     \loop \advance\epsfrsize\epsfrsize \divide\epsftmp 2
     \ifnum\epsftmp>0
        \ifnum\epsfrsize<\epsftsize\else
           \advance\epsfrsize-\epsftsize \advance\epsfysize\epsftmp \fi
     \repeat
     \epsfrsize=0pt
    \else
     \epsfrsize=\epsfysize
    \fi
   \fi
   \ifepsfverbose\message{#1: width=\the\epsfxsize, height=\the\epsfysize}\fi
   \epsftmp=10\epsfxsize \divide\epsftmp\pspoints
   \vbox to\epsfysize{\vfil\hbox to\epsfxsize{%
      \ifnum\epsfrsize=0\relax
        \includegraphics{\ifepsfdraft}%
      \else
        \epsfrsize=10\epsfysize \divide\epsfrsize\pspoints
        \includegraphics{\ifepsfdraft}%
      \fi
      \hfil}}%
\global\epsfxsize=0pt\global\epsfysize=0pt}%
{\catcode`\%=12 \global\let\epsfpercent=
\long\def\epsfaux#1#2:#3\\{\ifx#1\epsfpercent
   \def\testit{#2}\ifx\testit\epsfbblit
      \epsfgrab #3 . . . \\%
      \epsffileokfalse
      \global\epsfbbfoundtrue
   \fi\else\ifx#1\par\else\epsffileokfalse\fi\fi}%
\def\epsfempty{}%
\def\epsfgrab #1 #2 #3 #4 #5\\{%
\global\def\epsfllx{#1}\ifx\epsfllx\epsfempty
      \epsfgrab #2 #3 #4 #5 .\\\else
   \global\def\epsflly{#2}%
   \global\def\epsfurx{#3}\global\def\epsfury{#4}\fi}%
\def\epsfsize#1#2{\epsfxsize}
\begin{document}
\title{ THE COLD PLUS HOT DARK MATTER MODEL FROM SUPERSYMMETRIC INFLATION}
\author{ Qaisar Shafi and Robert K. Schaefer }
\address{Bartol Research Institute, University of Delaware, \\ 
Newark, DE 19716, USA}
\maketitle
\abstracts{
The cold plus hot dark matter (CHDM) model is arguably the best theory
we currently have for a consistent description of the observed large 
scale structure formation. This is especially true if the primordial density
fluctuations are assumed to be essentially scale invariant, in which case
a mixture with 20-25 \% HDM, 5- 10 \% baryons, and the rest in CDM correctly
predicted (in 1989) the quadrupole anisotropy measured a few years later by the
COBE satellite. After a brief historical introduction, we present a model
of supersymmetric inflation in which the CHDM model is neatly realized
with a spectral index n= 0.98, while the dark matter consists of a few eV
`tau' neutrino and the LSP (essentially the `bino'). We also provide a
comparison of this model against the observations.
}
\section{ Introduction}
\label{intro}
   The idea that the dark matter in the universe may contain more than one
non--baryonic component does not seem so `strange' to this modern audience, 
but it was greeted with much skepticism when it was put forward in 1984 
\cite{ss84} as the basis
for a model of large scale structure formation. The inspiration for the CHDM
model came from both particle physics and cosmology. As people began to search
for models with axionic CDM, it quickly became clear that some of the best
models \cite{holman} also predicted non-zero neutrino masses.  More 
importantly, it was shown \cite{ss84,sss} that the presence of some amount of 
hot dark matter could go a
long way in reconciling the (critical density) inflationary scenario
with the observations. 

 Perhaps an analogy with particle physics can help highlight the situation.  
Even though the $SU(2)\times U(1)$ group may not appear as 
`attractive' as say just $SU(2)$ or $SU(3)$, the fact is that a description
of the electroweak interactions including both quarks and leptons could not 
be simulataneously achieved within the `nice' looking $SU(2)$ or $SU(3)$, (even
before the discovery of the weak neutral current).   
Similarly, you need
the CHDM (and not the CDM) model to provide a consistent description of 
observations on scales varying between the galactic and the horizon size. 
This holds especially if the primordial density fluctuations are assumed to 
be scale invariant, as was the case in the original CHDM model.
This model received a boost in 1992 when it became clear \cite{ssnature} 
that a model \cite{sss} with approximately 25 \% HDM, 5-10\% baryons, and the 
rest in CDM would beautifully match with the temperature fluctuation amplitude 
observed by the COBE satellite \cite{COBE1}.  It needs to be stressed that 
this was a prediction, NOT a `postdiction'!

Now to the rest of this talk. We first want to show how the CHDM model `fits'
in with the recent wave of interest in supersymmetry. In particular, we
present a very simple framework for realizing an inflationary scenario
with CHDM as the end product. Among other things, this model
\cite{dvali94,lss96} has a 
spectral index n=0.98, negligible `gravity' waves, and a density fluctuation 
amplitude which is proportional to $( M/M_{Planck} )^2$, where $M\approx
M_{GUT}$ denotes the gauge symmetry breaking scale.  We conclude by 
presenting a comparison of this model with the current, ongoing, and planned 
observations.

\section{Supersymmetric Inflation}

As we will see in section 3 (also see talks by Liddle and Primack in these
proceedings), the `standard critical density CHDM model' with $n\approx 1$
provides a good fit to the present data on large scale structure. Several
questions can now be asked: How can the CHDM model arise from an inflationary
framework?  Can inflation be associated with some gauge symmetry breaking in
the early universe?  What is the nature of the `cold' and `hot' components? 

  Since the standard $SU(3)\times SU(2)\times U(1)$ model has no obvious dark 
matter candidate, while its supersymmetric extension (also known as MSSM) only
contains CDM, it seems clear that we should search for an inflationary model
based on a larger gauge symmetry.  One simple framework is offered by the
supersymmetric extension of the left--right symmetric gauge models.  The
`light' neutrinos in this scheme are necessarily massive, and with the LSP as
`cold' dark matter, we have a simple particle physics basis for the CHDM model. 
Remarkably, we find that the $SU(3)_c \times SU(2)_L \times SU(2)_R \times 
U(1)_{B-L}$ gauge model not only admits inflation, but that we can use it 
to `pin down' the symmetry breaking scale of $SU(2)_R \times U(1)_{B-L}$.  
It turns out to be
comparable to the SUSY GUT scale ($\sim 10^{16}$ GeV).  The hope, of course, is
that such models can be embedded in a supersymmetric grand unified framework. 

Consider the following globally supersymmetric renormalizable
superpotential $W$ \cite{inflsup}:
\begin{equation}
W\ =\ \kappa S \bar{\phi} \phi \ -\ \mu^2 S\ \ (\kappa>0,\ \mu>0), 
\end{equation}
where $\phi,\ \bar{\phi}$ denote the standard model singlet components of a
conjugate
pair of $SU(2)_R \times U(1)_{B-L}$ doublet left handed superfields, and
$S$ is a gauge
singlet left handed superfield. An R-symmetry, under which $S \rightarrow
e^{i \alpha} S,\
\bar{\phi} \phi \rightarrow \bar{\phi} \phi$, and $W
\rightarrow e^{i \alpha} W$, can ensure that the rest of the renormalizable
terms are either absent or irrelevant. Note that the
gauge quantum numbers of $\phi$ are precisely those of the
`matter'
right handed neutrinos. But they are distinct (!) superfields and, in
particular, the latter do
not have the conjugate partners. From $W$, one writes down the potential
$V$ as a
function of the scalar fields $\phi, \bar{\phi}, S$:
\begin{eqnarray}
V(\phi, \bar{\phi}, S)\ &=& \ \kappa^2 \mid S \mid^2 \ [\ \mid \phi \mid^2  +
 \mid \bar{\phi}
\mid^2\ ] \nonumber \\ 
&+& \mid \kappa \phi \bar{\phi} - \mu^2 \mid^2  +  D-{\rm terms}.
\end{eqnarray}
The D-terms
vanish along the D-flat direction $\phi = \bar{\phi}^*$ which contains the
supersymmetric
minimum
\begin{equation}
\begin{array}{ccl}
\langle S\rangle & = & 0, \\
\\
\langle \mid \phi \mid \rangle & = & \langle \mid \bar{\phi} \mid \rangle =
\mu/\sqrt{\kappa}
\equiv M.
\end{array}
\end{equation}
Using an appropriate R-transformation, $S$ can be brought to the real axis,
i.e.,
$S=\sigma/\sqrt{2}$, where $\sigma$ is a normalized real scalar field.

The important point now is that in the early universe the scalar fields are
displaced from
the above minimum. In particular, for $S > S_c = M$, the potential $V$ is
minimized by
$\phi = \bar{\phi} = 0$. The energy density is dominated by $\mu^4$ which
therefore
leads to an exponentially expanding inflationary phase (hybrid inflation).
As emphasized
in \cite{dvali94}, there are important radiative corrections under these
conditions \cite{soft}. At one
loop, and for $S$ sufficiently larger than $S_c$, the inflationary
potential is given by
\begin{equation}
V_{eff} (S) = \mu^4 \left[ 1 + \frac{\kappa^2}{16\pi^2} \left( ln
\left(\frac{\kappa^2
S^2}{\Lambda^2}\right) + \frac{3}{2} - \frac{S_c^4}{12S^4} + \cdots \right)
\right].
\label{veff}
\end{equation}

Using equation (\ref{veff}), one readily finds \cite{liddle93} the fundamental
quantity: 
\begin{equation}
\label{quad}
(\Delta T/T)_Q \approx 8 \pi (N_Q/45)^{1/2} (M/M_{P})^2, 
\end{equation}
where $(\Delta T/T)_Q$ is the cosmic microwave quadrupole anisotropy
amplitude.  Here
$N_Q \approx 50 - 60$ denotes the relevant number of e-foldings experienced
by the
universe between the time the quadrupole scale exited the horizon and the
end of inflation.  We also find  the primordial density fluctuation spectral
index $n \simeq 0.98$. 
From equation (\ref{veff}), one finds $\kappa \approx \frac{8
\pi^{3/2}}{\sqrt{N_Q}} \ y_Q \left(\frac{M}{M_P} \right) $, where $y_Q=
x_Q(1-7/(12x_Q^2)
+\cdots)$ with $x_Q = S_Q/M$, and $S_Q$ is the value of the scalar
field $S$ when the scale which evolved to the present horizon size crossed
outside the
de Sitter horizon during inflation.

The inflationary phase ends as $S$ approaches $S_c$ from above. Write $S =
xS_c$,
where $x=1$ corresponds to the phase transition from $G \rightarrow H$
which,  it
turns out, more or less coincides with the end of the inflationary phase
(this is checked by
noting the amplitude of the quantities
$\epsilon = \frac{M_P^2}{16 \pi} (V^{\prime}/V)^2$ and $\eta =
\frac{M_P^2}{8 \pi} (V^{\prime
\prime}/V)$, where the prime refers to derivatives with respect to the
field $\sigma$). Indeed, the
$50-60$ e-foldings needed for the inflationary scenario can be realized
even with $x
\approx 2$. An important consequence of this is that with $S \sim 10^{16}$
GeV, the supergravity corrections are negligible \cite{sugr}.

In order to estimate the `reheat' temperature we take account of the fact
that the inflaton
consists of the two complex scalar fields $S$ and $\theta=(\delta \phi + \delta
\bar{\phi})/\sqrt{2}$, where $\delta \phi = \phi - M$, $\delta \bar{\phi} =
\bar{\phi} - M$, with
mass $m_{infl} = \sqrt{2}\kappa M$. We mainly concentrate on the decay
of $\theta$. Its relevant coupling to `matter' is provided by the
non-renormalizable
superpotential coupling (in symbolic form): \begin{equation}
\frac{1}{2}\left( \frac{M_{\nu^c}}{M^2} \right) \bar{\phi} \bar{\phi} \nu^c
\nu^c,
\end{equation}
where $M_{\nu^c}$ denotes the Majorana mass of the relevant right handed
neutrino
$\nu^c$. Without loss of generality we assume that the Majorana mass matrix
of the right
handed neutrinos has been brought to diagonal form with positive entries.
Clearly,
$\theta$ decays predominantly into the heaviest right handed neutrino
permitted by
phase space. (The field $S$ can rapidly decay into higgsinos through the
renormalizable
superpotential term $\xi S h^{(1)} h^{(2)}$ allowed by the gauge symmetry,
where $h^{(1)},\
h^{(2)}$ denote the electroweak higgs doublets which couple to the up and
down type
quarks respectively, and $\xi$ is a suitable coupling constant. Note that
after supersymmetry
breaking, $\langle S\rangle\sim M_S$, where $M_S\sim$ TeV denotes the
magnitude of the
breaking.)

Following standard procedures (we will soon comment on the issue of parametric
resonance), and assuming the MSSM spectrum, the `reheat' temperature $T_R$ is
given by
\begin{equation}
T_R\ \approx \frac{1}{7} \left( \Gamma_\theta M_P\right)^{1/2}, \label{reheat1}
\end{equation}
where $\Gamma_\theta \approx (1/ 16\pi) (\sqrt{2} M_{\nu^c}/M)^2 \sqrt{2}
\kappa M$ is the decay rate of $\theta$. Substituting $\kappa$ as a function 
of $N_Q$, $y_Q$, and $M$, we find
\begin{equation}
T_R \approx \ {1\over 12} \left( \frac{56}{N_Q} \right)^{1/4} \sqrt{y_Q}\
M_{\nu^c}.
\label{reheat2}
\end{equation}

Several comments are in order:
\begin{list}%
\setlength{\rightmargin=0cm}{\leftmargin=0cm}
\item[{\bf i.}] For $x_Q$ on the order of unity the `reheat' temperature is
essentially determined
by the mass of the heaviest right handed neutrino the inflaton can decay into;

\item[{\bf ii.}] The well known gravitino problem requires that $T_R$ lie
below $10^8-10^{10}$
GeV, unless a source of late stage entropy production is available. Given the
uncertainties, we will interpret the gravitino constraint as the
requirement that $T_R
\stackrel{_<}{_\sim} 10^{9}$ GeV.

\item[{\bf iii.}] In deriving equation (\ref{reheat2}) we have ignored the
phenomenon of parametric resonance. This is justified because the 
oscillation amplitude is of order $M$ (not $M_P$!), such that the induced
scalar mass ($\sim M_{\nu^c}$) is smaller than the inflaton mass
$\sqrt{2} \kappa M$.  Note that here $M_{\nu^c}$ denotes the mass of the
heaviest right handed neutrino super-multiplet the inflaton can decay into.  
\end{list}

To proceed further we will need some details from the see-saw mechanism for the
generation of light neutrino masses. For simplicity, we will ignore the
first family of quarks
and leptons. The Majorana mass matrix of the right handed neutrinos can then be
brought (by an appropriate unitary transformation on the right handed
neutrinos) to the
diagonal form with real positive entries \begin{equation}
{\cal M} = \left(\begin{array}{cc} M_1 & 0 \\ 0& M_2 \end{array} \right)\ \
\ (M_1,\ M_2>0).
\label{Mmatrix}
\end{equation}
An appropriate unitary rotation can then be further performed on the left
handed
neutrinos so that the
(approximate) see-saw light neutrino mass matrix $m_D {\cal M}^{-1}
\tilde{m}_D$, $m_D$
being the neutrino Dirac matrix, takes the diagonal form \begin{equation}
m_D\frac{1}{\cal M}\tilde{m}_D = \left(\begin{array}{cc} m_1 & 0 \\ 0 & m_2
\end{array}
\right).
\end{equation}
($m_1,\ m_2$ are, in general, complex) \cite{buchm90}. In this basis of
right and left
handed neutrinos, the elements of
\begin{equation}
m_D = \left(\begin{array}{cc} a & b \\ c& d \end{array} \right),
\label{diracmatrix}
\end{equation}
are not all independent. They can be expressed in terms of only three complex
parameters $a,\ d$, and $\eta$, where $\eta = - [M_1/M_2]^{1/2}(b/a) = [M_2
/M_1]^{1/2}(c/d)$. 

We will now assume that $m_D$ coincides asymptotically (at the SUSY GUT scale
$M_{GUT} \simeq 2\times 10^{16}$ GeV) with the up type quark mass matrix as
is the
case in many GUT models. 
Restricting ourselves, from now on, to the case
where $|\eta|\sim 1$ and $M_1/M_2 \gg 1$, we have $|a|\gg |b|$ and $|c| \gg
|d|$. Without
much loss of generality we can further take $|c| \ll |a|$ so that $a$ is
the dominant
element in $m_D$. Under these assumptions the asymptotic top and charm masses 
are $|m_t| \approx |a|$ and $|m_c| \approx |d|\ |1+\eta^2|$. Since 
$|m_2|\ll |m_1|$, we can make the following identification of the light 
neutrino mass eigenstates
\begin{equation}
m_{\nu_\tau} = |m_1|={|a|^2 \over M_1}|1+\eta^2|,\ \ m_{\nu_\mu} = |m_2|=
{|d|^2\over M_2}|1+\eta^2|. \label{mnus}
\end{equation}
We can then get the useful relations
\begin{equation}
M_2 \approx {m_c^2 m_t^2\over m_{\nu_\mu} m_{\nu_\tau}} {1\over M_1},\ \ \ \
|1+\eta^2|\approx {m_{\nu_\tau}\over m_t^2} M_1. \label{mscales}
\end{equation}

We are now ready to draw some important conclusions concerning neutrino
masses that
are more or less model independent. Assuming that the inflaton
predominantly decays to
the heaviest right handed neutrino ({\it i.e.} $M_{\nu^c}= M_1$ in equation
(\ref{reheat2}))
and employing condition (ii), we obtain $M_1 \stackrel{_<}{_\sim}
9.3\times 10^9$ GeV
for $N_Q\approx 56$ and $x_Q\approx 2$. Equation (\ref{mnus})
then implies 
an unacceptably large $m_{\nu_\tau}$ for $|\eta|\sim 1$. Thus, we are led
to our first
important conclusion: the inflaton should decay to the second heaviest
right handed
neutrino and consequently $M_{\nu^c} = M_2$ in equation (\ref{reheat2}).
Combining
this equation with equation (\ref{mscales}) we obtain
\begin{equation}
T_R \approx {1\over 12.1}\left({56\over N_Q}\right)^{1/4} {m_c^2 m_t^2\over
m_{\nu_\mu}
m_{\nu_\tau}} {y_Q^{1/2} \over M_1} \approx 1.2 \times 10^{22}\ {y_Q^{1/2}
\over M_1}\
{\rm GeV.}
\label{reheat3}
\end{equation}
Here we put $N_Q=56$ which is easily justifiable by standard methods at the
end of the
calculation after having fixed the values of all relevant parameters. Also,
we took $m_t =
120$ GeV, $m_c=0.25$ GeV, which are consistent with the assumption that below
$M_{GUT}$ the theory reduces to MSSM with large $\tan\beta$~\cite{babu}.
Moreover, we took $m_{\nu_\mu} \approx 10^{-2.8}$ eV which
lies at the center of the region consistent with the resolution of the
neutrino solar puzzle
via the small angle MSW mechanism. The value $m_{\nu_\tau} \approx 4$ eV  is
consistent with the light tau neutrino playing an essential role in the
formation of large scale structure in the universe. 

The value of $M_1$ is restricted by the fact that the inflaton should not
decay to the
corresponding right handed `tau' neutrino
\begin{eqnarray}
M_1 \geq {m_{infl}\over 2}= {\kappa M\over \sqrt{2}} &\approx& \left( {45
\pi \over
2}\right)^{1/2} {y_Q \over N_Q} M_P \left( {\Delta T\over T} \right)_Q
\nonumber \\
&\approx & y_Q\ 1.2\times 10^{13}\ {\rm GeV}. \label{bigm1}
\end{eqnarray}
It is interesting to note that since the right handed neutrinos acquire
their masses from
superpotential terms $\lambda \frac{\bar{\phi}\bar{\phi} \nu^c\nu^c}{M_c}$,
where
$M_c=M_P/\sqrt{8\pi}\approx 2.4\times 10^{18}$ GeV and $\lambda
\stackrel{_<}{_\sim}
1$, $M_1=2\lambda M^2/M_c \stackrel{_<}{_\sim} 2.9 \times 10^{13}$ GeV
($M\approx
5.9 \times 10^{15}$ GeV for $N_Q=56$, $(\Delta T/T)_Q = 6.6\times
10^{-6}$). Thus,
from equation (\ref{bigm1}), $y_Q \stackrel{_<}{_\sim} 2.4$ which implies $x_Q
\stackrel{_<}{_\sim} 2.6$, and restricts the relevant part of inflation at
values of $S\sim 10^{16}$ GeV. 

To maximize the primordial lepton asymmetry (see below) we choose the bound in
equation (\ref{bigm1}) to be saturated. Equation (\ref{reheat3}) then gives
\begin{eqnarray}
T_R &\approx& y_Q^{-1/2}\ 9.7 \times 10^8\ \left( {\Delta T/T \over
6.6\times 10^{-6} }
\right)^{-1} \left( {N_Q \over 56 } \right)^{3/4}\nonumber \\
& &\left( {m_c \over 0.25 {\rm GeV} } {m_t \over 120 {\rm GeV} } \right)^2 
\left( {m_{\nu_\mu} \over 10^{-2.8} {\rm eV} } {m_{\nu_\tau} \over 4 {\rm eV} }
\right)^{-1} {\rm GeV},
\label{reheat4}
\end{eqnarray}
which satisfies condition (ii) for all allowed values of $y_Q$. Eq.
(\ref{mscales}) implies
\begin{equation}
\label{m2eta}
M_2 \approx y_Q^{-1}\ 1.2\times 10^{10}\ {\rm GeV},\ \ |1 + \eta^2|
\approx 3.4\ y_Q.
\label{}
\end{equation}
This implies that the errors in the asymptotic formulas for the top and charm 
masses are $<1$\%.  

The observed baryon asymmetry of the universe can be generated by first
producing a
primordial lepton asymmetry via the out-of-equilibrium decay of the right
handed
neutrinos, which emerge as decay products of the inflaton field at `reheating'
\cite{fugu86}.  It is important though to ensure that the
lepton asymmetry is not erased by lepton number violating 2-2 scatterings
at all
temperatures between $T_R$ and 100 GeV \cite{harvey90}. In our case this
requirement
is automatically satisfied since at temperatures above $10^7$ GeV the
lepton asymmetry
is protected \cite{ibanez92} by
supersymmetry, whereas at temperatures between $10^7$ and 100 GeV, as one can
easily show, these 2-2 scatterings are well out of equilibrium. The out-of
-equilibrium
condition for the decay of the right handed neutrinos is also satisfied
since $M_2 \gg
T_R$ for all relevant values of $x_Q$. The primordial lepton asymmetry is
estimated to be \cite{fugu86} 
\begin{equation}
{n_L\over s} \approx {9\over 8 \pi} {T_R\over m_{infl}} {M_2 \over M_1} {
{\rm Im}
(m_D^\dagger m_D/ |\langle h^{(1)}\rangle|^2)^2_{21} \over (m_D^\dagger
m_D/ |\langle
h^{(1)}\rangle|^2)_{22} }. \end{equation}
Equation (\ref{diracmatrix}) combined with the fact
that $|c||d| \ll |a||b|$ then gives 
\begin{equation}
{n_L\over s} \stackrel{_<}{_\sim}  {9\over 8 \pi} {T_R\over m_{infl}} 
{M_2 \over M_1} {m_t^2 \over |\langle h^{(1)}\rangle|^2}, 
\end{equation}
which, using equations (\ref{mscales}) - (\ref{m2eta}) and the fact that
$|\langle h^{(1)}\rangle|\approx 174$ GeV for large tan$\beta$, becomes
\begin{eqnarray}
{n_L \over s} &\stackrel{<}{_\sim}& y_Q^{-7/2}\ 6.6 \times 10^{-9} \left(
{\Delta T/T \over
6.6\times
10^{-6} } \right)^{-4} \left( {N_Q \over 56 } \right)^{15/4}\nonumber \\
& &\left( {m_c \over
0.25 {\rm GeV} } \right)^4 \left( {m_t \over 120 {\rm GeV} } \right)^6
\left( {m_{\nu_\mu} \over
10^{-2.8} {\rm eV} }\ {m_{\nu_\tau} \over 4 {\rm eV} }
\right)^{-2}.
\end{eqnarray}
For $x_Q\approx 2$ ($y_Q\approx 1.7$), this gives $n_L/s
\stackrel{_<}{_\sim} 10^{-9}$
which is large enough to account for the observed baryon asymmetry. Also
$T_R \approx 7 \times 10^8$ GeV, $M_1\approx 2\times 10^{13}$ GeV, 
$M_2\approx 7 \times 10^{9}$ GeV, and $m_{infl} \approx 4 \times 10^{13}$ GeV 
for the same value of $x_Q$.

In supersymmetric models the lightest supersymmetric particle (LSP) is
expected to be stable and is a leading cold dark matter candidate. If we 
couple this with a tau neutrino of mass $\sim 2-6$ eV we are led to the well 
tested (CHDM) model. 

To summarize, among the key features of the inflationary models we have
discussed one could list the role played by radiative corrections in the early
universe, the realization of inflation at scales well below $M_P$ so that the
gravitational corrections can be adequately suppressed, and the constraints on
the two heaviest right handed neutrino masses.  The cold plus hot 
dark matter combination which results is an important consequence.  

\section{Comparison of predictions to Large Scale Structure Observations}
 
Inflationary critical density cold plus hot dark matter models were recently
tested against measurements of large scale structure (see ref. 
 [18]
and references therein). The model described here was shown to be quite
compatible with observations as long as the heaviest neutrino mass
$m_{\nu_\tau}\sim 2-7$ eV and the Hubble constant turns out to be $h
\stackrel{<}{_\sim} 0.55$, where $h$ is the value of the Hubble constant in
units of 100 km s$^{-1}$ Mpc$^{-1}$. It is interesting to note that the
restriction on the Hubble constant from the age of the universe (which was not
used in ref. 
[18]) yields a nearly identical constraint. 
Observational determinations of the Hubble constant have not yet settled down 
to a precise value.   Although some current determinations have found $h\simeq
0.7-0.8$, lower values are still being seen.  In fact, a recent survey 
\cite{tam96} of determinations using observations made by the Hubble Space 
Telescope finds $h=0.55\pm 0.10$.  However, the greatest difficulty with 
Hubble constant determinations is overcoming systematic errors, and we 
await with interest more precise results.  

\begin{figure}[t]
\centering
\leavevmode
\epsfxsize=8.0cm
\epsfysize=7.0cm \epsfbox{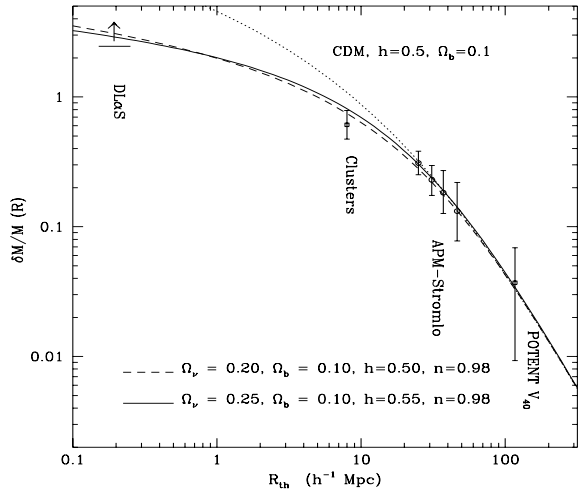}
\caption{We compare the predicted filtered density contrast 
(mass fluctuation amplitude $\delta M/M$) in this inflationary CHDM model
against observations.  We compare 
two $n=0.98$ models with $\Omega_{\nu} = 0.25$ and $h=0.55$ (solid line) and 
$\Omega_{\nu} = 0.20$ and $h=0.50$ against a variety of observations (see
text.)  We also show a CDM model ($\Omega_\nu=0$, $h=0.5$, and $n=1.00$) for 
comparison. The error bars are 2 $\sigma$ (95 \% confidence). 
}
\label{lssfig}
\end{figure}

Reference 
[18] assumed a baryon fraction of 
$\Omega_{baryon} =0.064\ (0.5/h)^2$, which is smaller than the currently 
allowed upper limits from big bang
nucleosynthesis.  The range consistent with
galactic chemical evolution and some recent high redshift deuterium
abundances, is \cite{hata96} 
\begin{equation}
\label{bbn}
0.08 \leq \Omega_b (h/0.5)^2 \leq 0.12, 
\end{equation}
which is somewhat higher.  This increase in the allowed baryon fraction helps
reconcile the high baryon fractions observed in clusters (see, e.g. 
ref. 
[21])  with a critical density CHDM universe.  Therefore we 
will assume that our critical density universe has a baryon fractions 
$\sim 10$ \%.  

   We now compare the supersymmetric inflationary model against observations.  
First of all, our model suggests that we have only one neutrino flavor with a
mass in the eV range, while another popular version \cite{caldwell} suggests 
that there are two nearly degenerate (in mass) flavors.  The reason is that 
with two flavors one can decrease the amplitude of cluster scale fluctuations 
while still getting a reasonable epoch of galaxy formation. 
However, as we have shown elsewhere\cite{babu95}, a larger
baryon fraction mimics the effect of increasing the number of neutrino flavors
in a lower baryon fraction model.  Thus we get comparable results to the two
flavor, lower baryon model.  

   The model has $\Omega=1$, $h<0.6$, $n=0.98$, 
$\Omega_b=0.10$, and 1 flavor of neutrino with mass in the few eV 
range.  To see how such a specific prescription compares with observations, we
plot in figure \ref{lssfig} the COBE normalized \cite{COBE} rms filtered 
density contrast as a function of the filtering length. 
The two models have $\Omega_\nu=0.25$, $h=0.55$ (solid) and 
$\Omega_\nu=0.20$, $h=0.50$ (dashed).  The observations are shown with 95\%
confidence limits, so a model with $\Omega_\nu=0$, $h=0.5$ (the CDM model -
dotted line) strongly violates observational limits.  The
observational limits in figure \ref{lssfig} are the large scale streaming 
velocities
(POTENT, ref. 
[24]), the galactic counts-in-cells measurements from
the APM - Mt. Stromlo survey\cite{apm} which have been corrected for a
linear bias factor of 1.4, the x-ray cluster abundance
constraint from ref. 
[18], and the 95\% confidence lower limit
required to make early galaxies (damped Ly-$\alpha$ systems - ref.
[26]).  

\begin{figure}
\centering
\leavevmode
\epsfxsize=8.0cm
\epsfysize=7.0cm \epsfbox{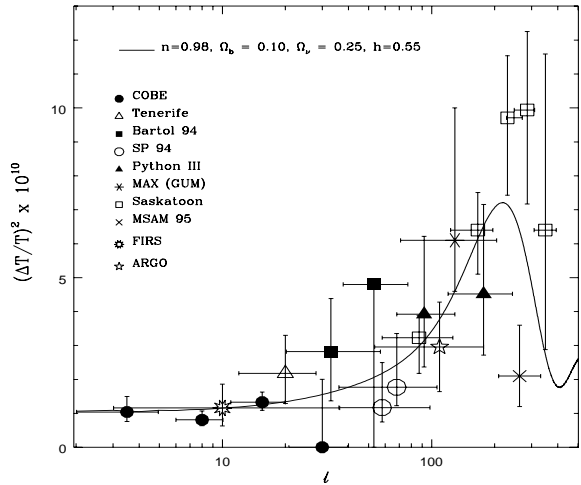}
\caption{We compare the predictions of temperature anisotropies in
this inflationary CHDM model against observations [29] (1 
$\sigma$ errors).   
The solid line is our inflationary model (supplied by A. A. de Laix). }
\label{cmb}
\end{figure}

   The cluster abundance is a synthesized linear theory constraint from x-ray 
measurements of clusters.  The best way to test these data are with detailed
hydrodynamic numerical simulations.  These are computer intensive and time
consuming, but allow for much more detailed comparison to observations.  We
note that simulations \cite{klypin} with similar parameters $n=1.00$, 
$\Omega_\nu = 0.2$, $h=0.5$, $\Omega_b= 7.5$ \% and two massive neutrino 
flavors show remarkably good agreement with observations. 

   The early galaxy formation constraint could be revised if many more high
redshift ($z>4$) galaxies are observed.  However, there are indications that
the galactic number density drops off beyond $z\sim 3$, indicating that most
galaxies formed relatively recently as we expect in a cold plus hot dark matter
universe (see, e.g. ref. 
[28] for a recent reference).  

   Lastly we compare the temperature anisotropy predictions of our models with
observations.  In figure \ref{cmb} we plot results from a variety of CMB 
experiments\cite{cmba} 
(1 $\sigma$ errors) along with predictions for our inflation model 
$\Omega_\nu=0.25$, $n=0.98$, $h=0.55$ and $\Omega_b=0.10$ (solid line)
We see that the present CMB data is consistent with the inflationary CHDM 
model although the data are not yet very discriminating.  Data from planned 
and ongoing experiments should be able to test this model much more 
precisely.  

  We see that starting from a very simple superpotential we have arrived not 
only at a successful model of inflation but also a beautiful picture of large
scale structure formation which is quite consistent with present large scale
structure observations.  For particle physics the most important predictions
include a massive `tau' neutrino in the 2-7 eV range, (a ``smoking gun" of the
CHDM model), as well as an LSP which
is more or less pure bino.  For cosmology, we predict an essentially scale
invariant spectrum (index $n\approx 0.98$) and an absence of gravity waves.  

\section*{Acknowledgments} One of us (Q.S.) would like to thank the organizers,
especially Hans V. Klapdor--Kleingrothaus and Yorck Ramachers for holding a
particularly stimulating conference covering many facets of `dark matter'
physics/astrophysics.  We also acknowledge fruitful collaborations with Gia
Dvali \& George Lazarides (on supersymmetric inflation and its ramifications)
and Andrew de Laix, Andrew Liddle, David Lyth, \& Floyd Stecker (CHDM model).


\begin{thebibliography}{99}
\bibitem{soft} The soft supersymmetry breaking scalar masses can be safely
ignored.
\end{thebibliography}
\end{document}